%% file: main.tex
\RequirePackage{rotating}
\documentclass[10pt, conference]{IEEEtran}

\input{macros}

\usepackage{graphicx}
\begin{document}

\title{Live Programming for Finite Model Finders}


\author{
\IEEEauthorblockN{Allison Sullivan}
\IEEEauthorblockA{
\textit{University of Texas at Arlington} \\
Arlington, TX USA \\
allison.sullivan@uta.edu}
}

%
%


%
%

%
%
\maketitle   

\begin{abstract}
Finite model finders give users the ability to specify properties of a system in mathematical logic and then automatically find concrete examples, called solutions, that satisfy the properties. These solutions are often viewed as a key benefit of model finders, as they create an exploratory environment for developers to engage with their model. In practice, users find less benefit from these solutions than expected. For years, researchers believed that the problem was that too many solutions are produced. However, a recent user study found that users actually prefer enumerating a broad set of solutions. Inspired by a recent user study on Alloy, a modeling language backed by a finite model finder, we believe that the issue is that solutions are too removed from the logical constraints that generate them to help users build an understanding of the constraints themselves. In this paper, we outline a proof-of-concept for live programming of Alloy models in which writing the model and exploring solutions are intertwined. We highlight how this development environment enables more productive feedback loops between the developer, the model and the solutions.
\end{abstract}

\input{intro}
\input{background}
\input{technique}
\input{nextsteps}

\input{relatedwork}
\input{conclusion}

%
%
%
\bibliographystyle{splncs04}
\bibliography{bib}

\end{document}

%% file: macros.tex
\usepackage{amsmath}
\usepackage[ruled,vlined,linesnumbered]{algorithm2e}
\usepackage{amsmath}
\usepackage[T1]{fontenc}
\usepackage{graphicx}
\usepackage{textcomp}
\usepackage{tikz}
\usepackage{hyperref}
\usepackage{textcomp}
\usepackage{color}
\usepackage[utf8]{inputenc}
\usepackage{fancyvrb}
\usepackage{xcolor}
\usepackage{colortbl}
\usepackage{hhline}
\usepackage{lscape}
\usepackage{upquote}
\usepackage{array}
\usepackage{adjustbox}
\usepackage{rotating}
\usepackage{stmaryrd}
\usepackage{pifont}
\usepackage{multirow}
\usepackage{wrapfig}
\usepackage{verbatim}
\usepackage{colortbl}
\usepackage{rotating}

\newcolumntype{L}[1]{>{\raggedright\arraybackslash}m{#1}}
\newcolumntype{C}[1]{>{\centering\arraybackslash}m{#1}}
\newcolumntype{R}[1]{>{\raggedleft\arraybackslash}m{#1}}

\fvset{commandchars=\\\{\}}

\definecolor{Green}{rgb}{0.0, 0.56, 0.0}
\definecolor{Gray}{gray}{0.85}
\definecolor{LightBlue}{cmyk}{0.51, 0.1, 0, 0}
\definecolor{Maroon}{cmyk}{0.32, 1.0, 1.0, .46}

\newcommand{\Blue}[1]{\textcolor{blue}{\textbf{#1}}}

\newcommand{\Red}[1]{\textcolor{red}{\textbf{#1}}}

\newcommand{\CodeIn}[1]{\begin{small}\texttt{#1}\end{small}}
\newcommand{\Comment}[1]{}
\newcommand{\NoType}[1]{} 
\newcommand{\Space}[1]{}

\newcommand{\DefMacro}[2]{\expandafter\newcommand\csname rmk-#1\endcsname{#2}}
\newcommand{\UseMacro}[1]{\csname rmk-#1\endcsname}

\DefMacro{check}{\ding{51}}
\DefMacro{corr}{\ding{72}}
\DefMacro{plausible}{\ding{51}}

\newcommand{\ACaret}{\raisebox{-0.6ex}{\textasciicircum}}
\newcommand{\AStar}{*}



%% file: intro.tex
\section{Introduction}

The expressibility of software modeling languages has greatly expanded and the automated analysis over these models has reached a point of efficiency where we can reason about real world systems~\cite{surgicalrobots,facebook,microsoft1,ChordAlloy,10.1007/s10009-019-00540-4,aws,aws2,nvidaalloy}. 
However, software models are not being widely adopted because there is still the bottleneck of writing the specification. While a formal methods expert can write the specifications, there is no guarantee that the expert will understand the domain area well enough to accurately capture the intricacies of the system. Ideally, the software architect would write the specifications. 

Unfortunately, modeling languages are notoriously difficult to learn, which is compounded by software modeling toolsets that lag behind the state-of-the-art for integrated development environments (IDE)~\cite{fmedu_tools,humanin_fm,siegel2021prototyping}. Most notably, modeling toolsets lack quality feedback. For instance, when a model is executed, the main result presented to the user is often a simple boolean result: either the analysis over the model is satisfiable or unsatisfiable. 
However, this does not give users enough context to answer the question ``did I write my model correctly?'' 
Additionally, a number of features that aide in composition, such as code completion, are almost universally absent. 

To address the feedback issue, there has been a rise in the creation of finite model finders, where users write a software model of their system's design and the finite model finder then produces concrete examples of behavior allowed by the model~\cite{JacksonAlloy2002,MACE,ProB,nitpick,forge,Miniatur,MemSAT,PBnJ}. In this paper, we will refer to the output of finite model finders as \textit{solutions} and the logical constraints that are executed as \textit{software models}. Clearly, in place of a boolean result, these solutions provide more context to the user about the behavior of their model. Furthermore, besides helping validate software designs \cite{CD2Alloy,Margrave,ChordAlloy,WickersonETAL2017,ChongETAL2018}, these solutions have been used to test and debug code~\cite{DiniETALKoratAPI2018,MarinovKhurshid01TestEra}, to repair program states~\cite{SamimiETALECOOP2010,ZaeemKhurshidECOOP2010} and to provide security analysis of systems~\cite{CheckMateMicro2019,websecurity,BagheriETAL2018}.  

While these solutions are often mentioned as one of the core benefits of finite model finders,  users have found them to be less helpful in practice~\cite{zave2015practical,alloyformalise23study,nelsonuserstudy}. A long held belief is that there are too many solutions, which can overwhelm the user. 
As a result, there is a whole body of work~\cite{Hawkeye,AUnit,PorncharoenwaseETALCompoSat2018,PonzioETALFSE2016,Seabs,Aluminum} dedicated to trying to reduce the number of solutions, often by creating additional criteria a solution must adhere too, such as minimality~\cite{Aluminum}. However, a recent user study found that users often abandon these tailored enumerations in favor of the default enumeration~\cite{nelsonuserstudy}, implying that users want more solutions to explore to better understand their model and thus their system. Furthermore, another user study with a mix of novice and expert users found that all users struggled with inspecting solutions and in turn, refining the model based on the solutions. 
Therefore, as a community we have been solving the \textit{wrong} problem.

\textit{Our theory} is that the solutions, while visually approachable, are too divorced from the logical constraints that form the model to actually help the user follow up on the question ``did I write my model correctly?'' Finite model finders do not convey \textit{why} solutions are generated, only that they \textit{are} generated. 
As a result, finite model finders place the burden on the user to determine how the abstract constraints they are writing ultimately influence the concrete solutions.
\textit{Our vision} is that a live programming environment, which interweaves writing the model and evaluating the model, is the answer to 
turning the solutions enumerated by finite model finders into constructive feedback, which in turn, can make software modeling approachable to the average software architect.

\input{figure/queue}



%% file: figure/queue.tex
\usetikzlibrary{shapes.geometric, arrows}
\tikzstyle{arrow} = [line width=1.5pt,->,>=stealth]
\tikzstyle{darrow} = [line width=1.5pt,<->,>=stealth]
\tikzstyle{NodeAtom} = [rectangle, minimum width=.75cm, minimum height=.75cm, text centered, draw=black, fill=yellow!75!orange]
\tikzstyle{ListAtom} = [rectangle, minimum width=.75cm, minimum height=.75cm, text centered, draw=black, fill=orange!75!yellow]

\begin{figure*}
    \centering

\begin{tabular}[t]{c||c|c||c|c}
\footnotesize (a) &  \multicolumn{2}{c||}{\footnotesize (b)}  &  \multicolumn{2}{c}{\footnotesize (c)} \\
\begin{minipage}[t]{.85\columnwidth}
\scriptsize
\begin{Verbatim}[]
 1. \Blue{one sig} Queue \{ \Blue{var} head : \Blue{lone} Node \}
 2. \Blue{var sig} Node \{ \Blue{var} link : \Blue{lone} Node \}
 3. \Blue{fact} WellFormed \{
 4.    \Blue{always all} n : Node | n !\Blue{in} n.^link
 5.    \Blue{always all} n : Node | n \Blue{in} Queue.head.*link
 6. \}
 7. \Blue{pred} dequeue \{
 8.    head' = head.link
 9.    \Blue{all} n : Queue.head.^link | n.link = n.link'
10. \}
11. \Blue{run} dequeue \Blue{for} \Red{3}
\end{Verbatim}
\end{minipage}

&

\begin{minipage}[t]{.3\columnwidth}
\begin{center}
\footnotesize
\begin{tikzpicture}[baseline,node distance=1.5cm]
\node (Q0) [ListAtom] {\textbf{Q0}};
\node (N0) [NodeAtom, below of=Q0] [align=center]{\textbf{N0}};
\node (N1) [NodeAtom, right of=N0] [align=center]{\textbf{N1}};
\begin{scope}[brown]
\draw [arrow] (N0) -- node[anchor=south] {\textbf{link}} (N1);
\end{scope}[brown]
\begin{scope}[red]
\draw [arrow] (Q0) -- node[anchor=west] {\textbf{head}} (N0);
\end{scope}[red]
\end{tikzpicture}
\end{center}
\end{minipage}

&

\begin{minipage}[t]{.15\columnwidth}
\begin{center}
\footnotesize
\begin{tikzpicture}[baseline,node distance=1.5cm]
\node (Q0) [ListAtom] {\textbf{Q0}};
\node (N0) [NodeAtom, below of=Q0] [align=center]{\textbf{N1}};
\begin{scope}[brown]
\end{scope}[brown]
\begin{scope}[red]
\draw [arrow] (Q0) -- node[anchor=west] {\textbf{head}} (N0);
\end{scope}[red]
\end{tikzpicture}
\end{center}
\end{minipage}

&

\begin{minipage}[t]{.15\columnwidth}
\begin{center}
\footnotesize
\begin{tikzpicture}[baseline,node distance=1.5cm]
\node (Q0) [ListAtom] {\textbf{Q0}};
\node (N0) [NodeAtom, below of=Q0] [align=center]{\textbf{N0}};
\begin{scope}[brown]
\end{scope}[brown]
\begin{scope}[red]
\draw [arrow] (Q0) -- node[anchor=west] {\textbf{head}} (N0);
\end{scope}[red]
\end{tikzpicture}
\end{center}
\end{minipage}

&

\begin{minipage}[t]{.15\columnwidth}
\begin{center}
\footnotesize
\begin{tikzpicture}[baseline,node distance=1.5cm]
\node (Q0) [ListAtom] {\textbf{Q0}};
\begin{scope}[brown]
\end{scope}[brown]
\begin{scope}[red]
\end{scope}[red]
\end{tikzpicture}
\end{center}
\end{minipage}

\\
& \footnotesize \textbf{(State 0)} & \footnotesize \textbf{(State 1)} & \footnotesize \textbf{(State 0)} & \footnotesize \textbf{(State 1)}

\end{tabular}
    \caption{Alloy Model of a Queue Data Structure}
    \label{fig:queue}
\end{figure*}
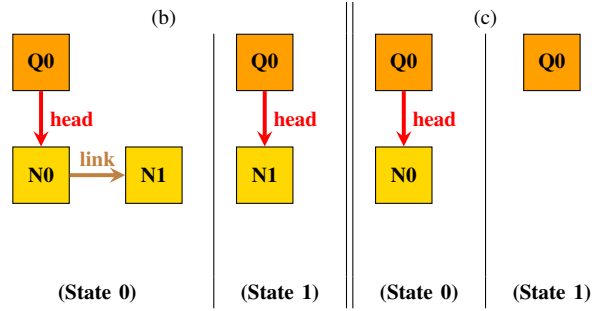

%% file: background.tex
\section{Motivating Example}\label{sec:example}
In this paper, we illustrate a live programming environment for the modeling language Alloy~\cite{JacksonAlloy2002}. Alloy enables the specification of both structural and behavioral properties in the form of relational, first order  and linear temporal logic constraints. Alloy is packaged in an automated analysis engine, the Analyzer, which utilizes Pardinus~\cite{Pardinus}, a temporal relational model finder, to enumerate solutions.

To highlight the limitations of the current model development process, consider the following model of a queue data structure seen in Figure~\ref{fig:queue} (a). To start, the user would write the model in the text editor portion of the Analyzer. Signature blocks introduce atoms and their relationships (lines 1 - 2). The keyword \CodeIn{var} indicates which portions of the model can change between states. 
For example, line 1 introduces a named set \CodeIn{Queue} and defines a mutable relation \CodeIn{head} that expresses that each \CodeIn{Queue} atom has zero or one (\CodeIn{lone}) node at the start of the queue. 
Users can write named formulas in fact (\CodeIn{fact}) blocks, which have to be true for any solution found, or predicate (\CodeIn{pred}) blocks, which have to be true for a solution if the predicate is invoked. To outline well-formed queues, the user specifies no node is reachable from itself (line 4) and all nodes are in the queue (line 5). To outline \CodeIn{dequeue}, the user specifies 
the new head of the queue is the second node in the queue (line 8) and there are no other changes to the order of the nodes in the queue (line 9). 

When the user is ready to check their constraints, the user executes the run command on line 11. The Analyzer will then use Pardinus to 
search for all satisfying solutions such that \CodeIn{dequeue} and \CodeIn{WellFormed} are true using up to 3 \CodeIn{Queue} atoms and 3 \CodeIn{Node} atoms. 
These solutions are rendered graphically in a separate pop up window, where the user can iterate over the solutions one by one. To illustrate, Figure~\ref{fig:queue} (b) and (c) display the first two  solutions the Analyzer finds. If the user encounters a malformed solution, the user now knows that their model allows for behavior she intended to prevent. The user then returns to the text editor and tries to update their current constraints to prevent the malformed solution. However, as the user makes edits, the only way to know the impact on the solutions is to re-execute the command and re-start the enumeration and inspection process. 

Ultimately, in this workflow, the burden is on the user to to mentally visualize the impact abstract constraints will have on the concrete shape of the solutions,
as editing the model and enumerating solutions are distinct tasks. However, if we could integrate writing constraints with the construction of the solutions users could witness the impact of their constraints in real time.


%% file: technique.tex
\section{Live Programming for Finite Model Finders}\label{sec:technique}

\begin{figure}
\centering
\includegraphics[width=8cm]{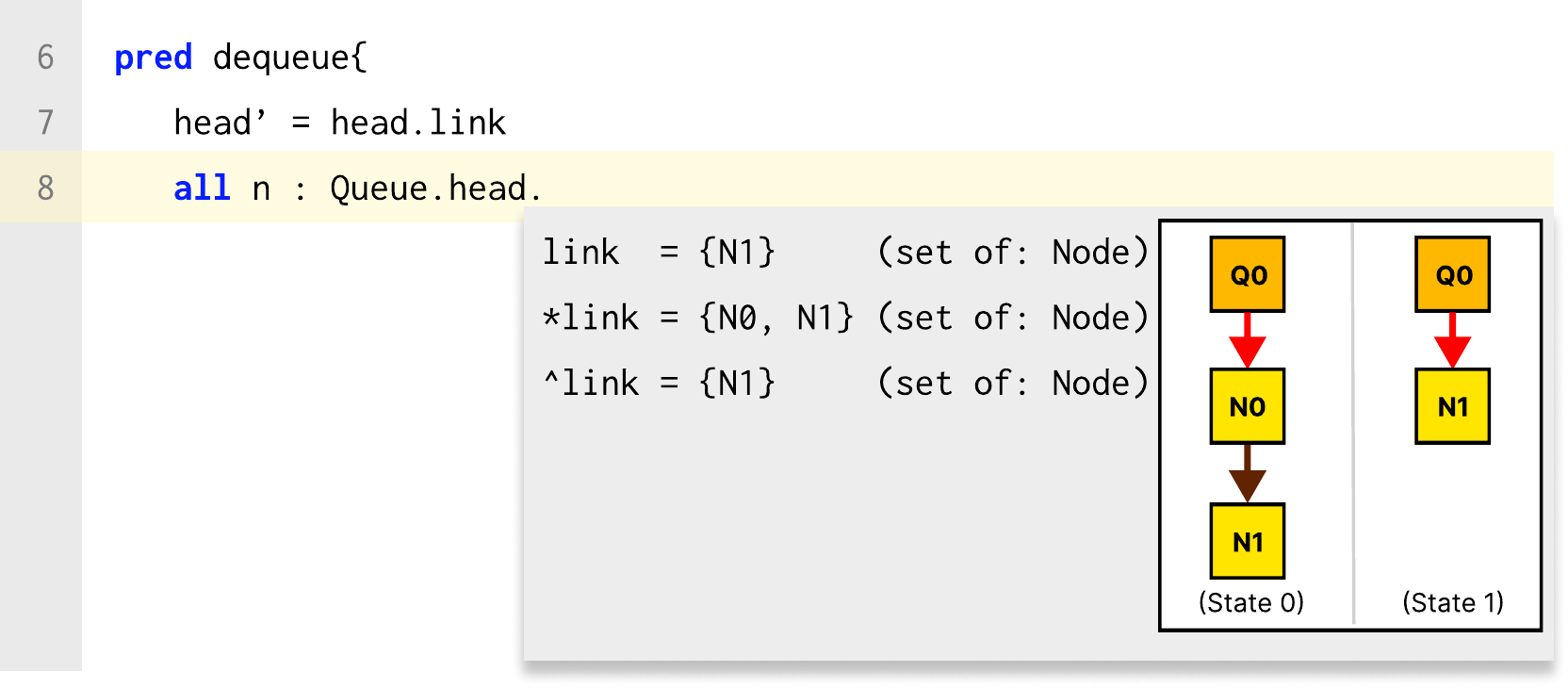} 
\caption{Prototype of Suggestion Box for Formula Completion}
\label{fig:projection}
\end{figure}

In this section, we outline our proof-of-concept of a live programming development environment for Alloy. First, we introduce suggestion boxes that provide formula completion suggestions with concrete illustrations of how each of the suggested formulas would behave. Second, we highlight two different development views that a user can toggle between: (1) a dynamic Enumeration View, which presents the broad impact across all possible solutions and (2) a Focus View, which presents the narrow impact on a single solution

\subsection{Formula Completion Suggestions}
Code completion, such as presenting valid API calls to make on an object, is a common feature in many integrated development environments (IDE) that is a lightweight intervention to help users efficiently compose their programs. The equivalent concept for a modeling language would be formula completion where we distill for the user valid extensions of the formula they are actively writing. We have begun developing a series of rules for formula completion based on using grammar and type rules to produce valid extensions. For instance, a relational join ``\CodeIn{a.b}'' will default to the empty set if there is a type mismatch between \CodeIn{a} and \CodeIn{b}, e.g. \CodeIn{link.head} which joins types \CodeIn{(Node$\times$\underline{Node}).(\underline{List}$\times$Node)} will always be empty while \CodeIn{head.link}, which joins types \CodeIn{(List$\times$\underline{Node}).(\underline{Node}$\times$Node)} will not. Therefore, if the user starts typing ``\CodeIn{link.}'' we do not want to suggest \CodeIn{head} as a continuation, but we would want to suggest \CodeIn{link} if the user is typing ``\CodeIn{head.}''. 
In addition, within Alloy we have a lightweight mechanic for constraint checking through the Evaluator, a toolset carried over from KodKod~\cite{KodKod}. This allows us to provide more context by annotating our suggestions with their actual value over a solution of the user's choosing. We combine completion suggestions with projected values into the concept of a suggestion box.

\begin{figure}
\centering
\includegraphics[width=8cm]{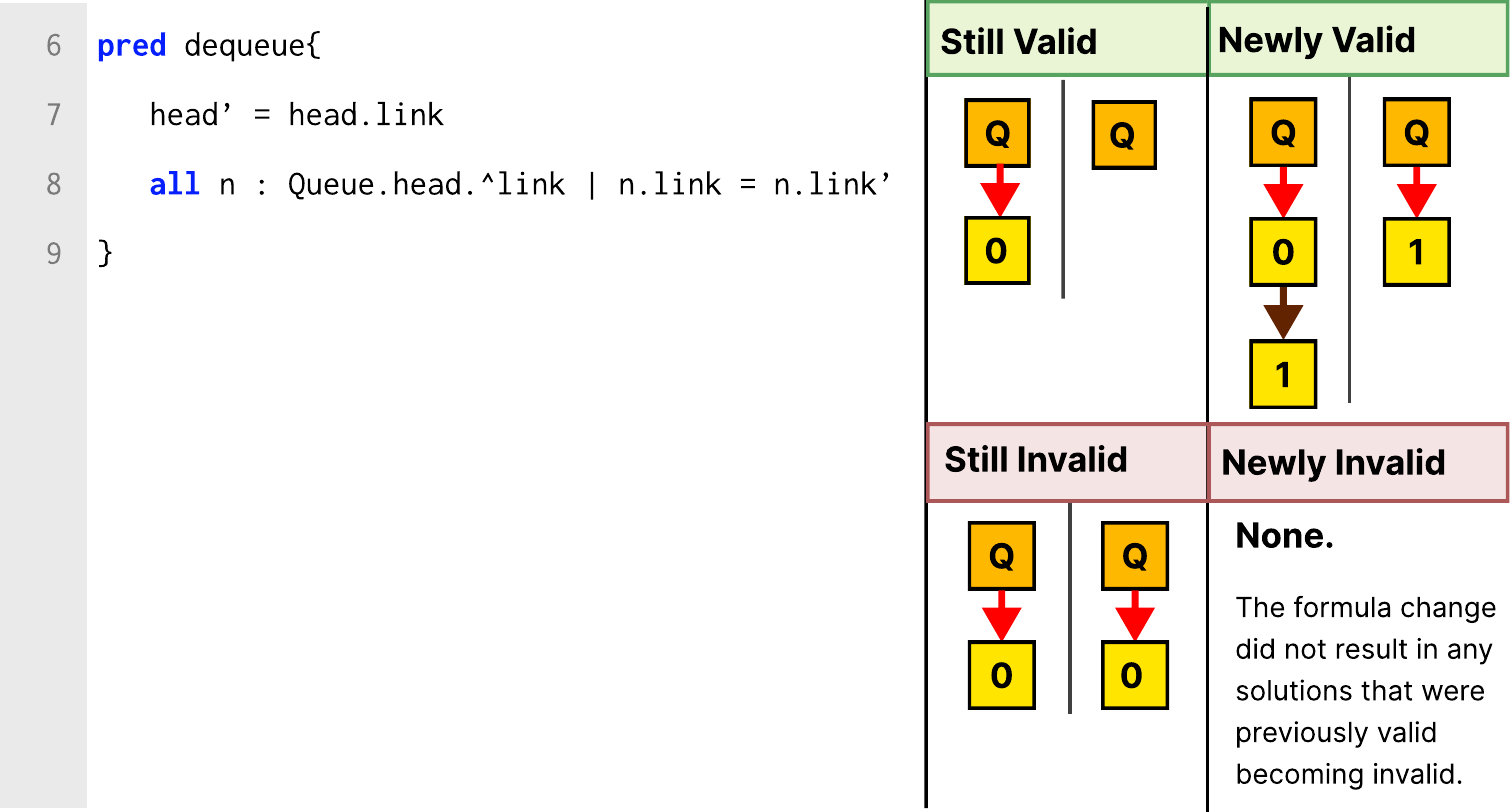} 
\caption{Prototype of Enumeration View}
\label{fig:enumeration}
\end{figure}

To illustrate, consider the user trying to capture ``all nodes in the queue except for the first node'' -- which is the domain of the quantified formula from \CodeIn{dequeue}. 
On their own, the user may struggle to determine the order to place the relational joins and whether to use reflexive transitive closure (`\CodeIn{\AStar}') or transitive closure (`\CodeIn{\ACaret}'). Figure~\ref{fig:projection} highlights how our suggestion box can aid the user in composing this domain. As the user types ``\CodeIn{Queue.head\Red{.}},'' the only built in set from the base model that does not produce an empty set is \CodeIn{link}. Therefore, the suggestion box populates with \CodeIn{link} and common extensions of \CodeIn{link}. With the added context of the value of these extensions over a solution, the user can determine that ``\CodeIn{\ACaret{link}}'' likely matches their intention.

In our final live programming IDE, we plan to enable the user to swap the solution, which will update the projected values. For instance, the user could update the solution in Figure~\ref{fig:projection} to a queue with 3 nodes, and the difference between selecting ``\CodeIn{link}'' and ``\CodeIn{\ACaret{link}}'' would be revealed. In addition, we plan to explore how to make more complex suggestions by generating our own set of common formula templates. RexGen, a relational algebra expression generator~\cite{AGenABZ}, can produce formulas to further populate our suggestion boxes. However, in practice, RexGen produces too many formulas to use as is. 
Therefore, we plan to use a recently published corpus of all Alloy models on Github~\cite{staticprofilealloy} to distill common templates for formula structures, which we will combine with RexGen to produce a small collection of complex suggestions. These common formula templates can also be utilized for other research, including the new but active field of automated repair for Alloy~\cite{ARepair,beafix,tar,icebar,atr}.


\begin{figure*}
\centering
\includegraphics[width=17cm]{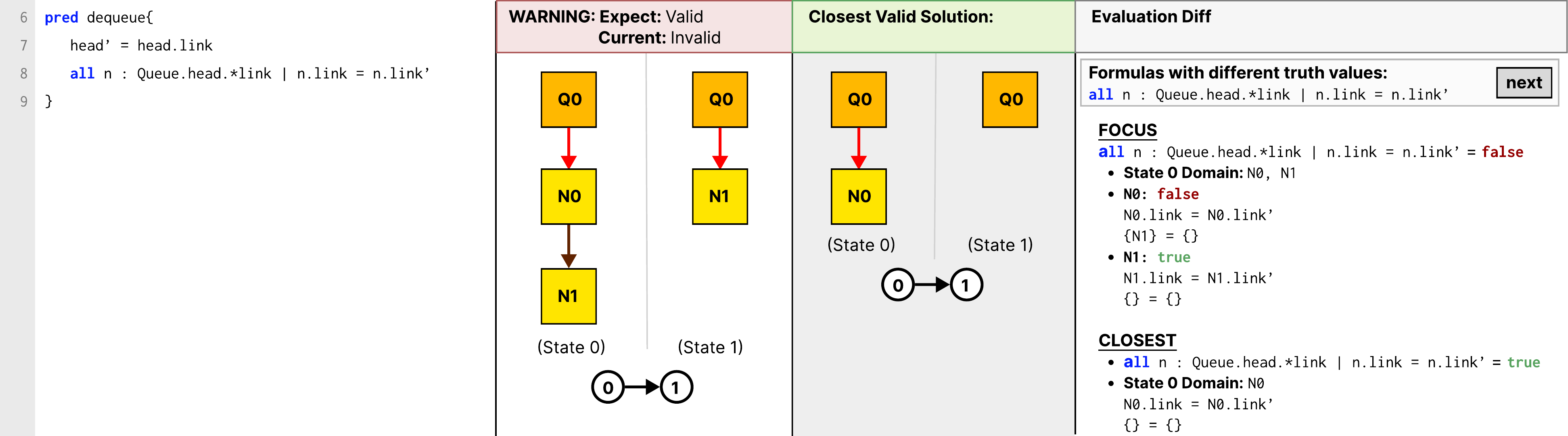} 
\caption{Prototype of Focus View}
\label{fig:focus}
\end{figure*}


\subsection{Enumeration View}
The main goal of live programming is for users to get instant feedback while changing their program. For an Alloy model, this feedback is the collection of \textit{all} possible solutions enabled by their model. 
Therefore, as the user adds constraints to the model, the user is constantly shifting the boundary of valid and invalid solutions. 
To bring this changing boundary to the forefront, our Enumeration View actively display two sets of solutions side by side with the text editor that update every time the user's edits create a model that compiles: a set of valid and a set of invalid solutions. This is inspired by a recent user study that found that novice users better understood pre-written Alloy constraints when presented with a combination of valid and invalid solutions~\cite{negpos}. The user study, while a promising result, used hand selected solutions. Our live programming environment will instead generate these solutions automatically. Since we are working with logical constraints, we can go beyond simply asking ``what solutions are valid?'' and ``what solutions are invalid?'' Instead, we can present four categories of solutions to the user: two valid categories -- (a) solutions that remained valid and (b) solutions that are now valid, and two invalid categories -- (c) solutions that stayed invalid and (d) solutions that became invalid. This is possible since we can use Alloy itself to compare and contrast two formulas. For instance, for two iterations of a formula (\CodeIn{A} and \CodeIn{A'}), we can execute the following command to get category (a) ``\CodeIn{\Blue{run} \{A \Blue{and} A'\}}'' and ``\CodeIn{\Blue{run} \{!A \Blue{and} A'\}}'' to get category (b).

To illustrate, Figure~\ref{fig:enumeration} shows the Enumeration View that would be produced if the user went from a faulty \CodeIn{dequeue} predicate in which the user used reflexive transitive closure to specify the domain of the quantified formula (``\CodeIn{Quene.head.\AStar{link}''} to the correct \CodeIn{dequeue} predicate in Figure~\ref{fig:queue} (a). Since each category would be maintained by a command, the user can enumerate solutions within each category. However, to avoid too high of a computational overhead, we will first use constraint checking to see if the currently displayed solution for a given category is still representative. If not, then we will generate a new solution. 

\subsection{Focus View}
While the output of a model is the collection of all solutions, as a user builds a model, it is not uncommon for the user to have a handful of key solutions in mind that the user is expecting to confirm that their model should generate or prevent. In fact, the adhoc practice of outlining a solution within a predicate to reason over individually was distilled into a unit testing framework~\cite{AUnit}. Leaning into this practice, the Focus View allows the user to see the impact on a single solution as they write their constraints. 
Within the Focus View, the user selects a solution and labels whether the solution is expected to be valid or invalid for their model. As the user writes, we constantly display the solution and it's current behavior (valid or invalid). 
As a result, the user is actively aware of any changes that result in discrepancies.

Furthermore, we use Alloy to provide debugging information to aid the user in understanding the current boundary between the solution the user expects to be valid (or invalid) and the behavior the current model prevents (or enables). The debugging information comes in two forms. First, for a solution $s$ whose behavior violates its expectation, we present the closest valid (or invalid) solution to $s$. 
We can generate the closest behavioral solution using a Partial-Max-SAT solver. For instance, for a solution that is expected to be valid but is not, we can make the hard constraints those enforced by the model and the soft constraints those outlining $s$. To provide even more context, we can also automatically decompose the difference between $s$ and the closest behavioral solution by presenting a breakdown of all the formulas in the model in which the two solutions produce different truth values. This evaluation breakdown can be done efficiently by turning the Evaluator from a black-box toolset into a white-box toolset that presents the intermediate evaluations discovered along the way to producing the final truth result.

To illustrate, Figure~\ref{fig:focus} shows the Focus View in action. The first pane is the text editor where the user just added the faulty quantified formula for \CodeIn{dequeue} with the incorrect domain mentioned in the Enumeration View example. In the second pane, the user has the solution from Figure~\ref{fig:queue} (b) in view. This solution was valid when the user only had the formula on line 7 written; however, after adding the formula on line 8, the solution is now prevented.
Therefore, the third pane displays the current closest valid solution, which reveals to the user that their formula results in valid behavior when the queue starts with one node but not when the queue starts with two nodes.  To further contextualize this, the final pane shows a breakdown of the faulty quantified formula across the two solutions. This breakdown reveals to the user that their current model incorrectly includes the first node in the queue within the quantified domain. 

%% file: nextsteps.tex
\section{Future Plans}
To bring our live programming environment from a proof-of-concept to a reality, there are two high level problems to address throughout. First, while we have initial prototypes for our different interfaces, we will need to refine these designs to ensure a smooth transfer of knowledge without overwhelming the user with too much information. In addition to how we present information, there is also a question of what information is best to present. For instance, for the starting solutions in our Enumeration View -- Should the solutions be as close to one another as possible across categories? Should we present maximal solutions to convey more context or is this too cluttered? Given the usability focus of this work, it is important that these decisions be vetted through active engagement with novice and expert end users. To that end, we plan to conduct multiple user studies as we build out our live programming framework. We anticipate using students as novice users and active members of the Alloy discourse group as expert users. 

Second, we need to produce responsive implementations of these interfaces. 
Across our different live programming designs, we often take advantage of Alloy itself to add rich information, such the value of suggestions or generating boundary solutions. However, Alloy's runtime is not nominal.
Therefore, we need to utilize existing optimizations and develop our own to ensure that the tandem presentation of solutions does not make the development environment sluggish. For instance, our final backend implementation of the Enumeration View will need to be carefully designed. There are existing bodies of work for incremental analysis of Alloy models, which can help ease the runtime overhead when we need to explicitly search for new solutions~\cite{titanium,platinum,Reach,iAlloy}. In addition, we can look for opportunities to use constraint checking to determine the impact of changes in order to delay, or even avoid, invoking a SAT solver for all interfaces. Along the way, we also expect to port the Analyzer codebase from Java to facilitate better implementations of these visual interfaces. 

%% file: relatedwork.tex
\section{Related Work}
Live programming is an active research field~\cite{tanimoto2013perspective,mcdirmid2007living,kubelka2018road,mcdirmid2013usable,10.1145/215585.215636,black2010pharo,deline2015tempe,edwards2005subtext,wilcox1997does,omar2019live,burnett1998implementing,ingalls2008lively,omar2021filling} that has been applied to variety of imperative, declarative and functional languages. To the best of our knowledge, live programming for modeling languages has only been explored for finite state machines~\cite{tikhonova2018constraint,van2019toward}. Our suggestion boxes are in the same spirit as projection boxes for Python, which display the live value of variables~\cite{lerner2020projection}. 
The research most closely related to our work for Alloy is (1) Amalgam~\cite{Amalgam}, which uses provenance chains to inform the user why a specific tuple does or does not appear in a solution and (2) abstract instances~\cite{abstractalloy}, which decomposes a solution into the parts present to satisfy the facts versus the parts present to satisfy predicates. Both of these aim to help the user understand why a solution was generated, and are complementary work that could be incorporated into a live programming environment to provide on-the-fly explanations of solutions.

%% file: conclusion.tex
\section{Conclusion}
Currently, the disconnect between the logical constraints that form a model and the graphical solutions that get produced by the model prevent users from utilizing solutions to better understand the logical constraints themselves. This paper presents the concept of a live programming environment for Alloy that interweaves writing and evaluating a model. Our proof-of-concept highlights how the interconnection of these two artifacts enables users to actually address the question ``did I write my model correctly?'' 